\documentstyle[aps,preprint]{revtex}
\begin{document}
\draft
\tightenlines
 
\title{First- and second-order phase transitions in a driven lattice gas with 
nearest-neighbor exclusion}
\author{Ronald Dickman$^{*}$}
\address{
Departamento de F\'{\i}sica, ICEx,
Universidade Federal de Minas Gerais,\\
30123-970
Belo Horizonte - MG, Brasil\\
}
\date{\today}

\maketitle
\begin{abstract}
A lattice gas with infinite repulsion between particles separated by 
$\leq 1$ lattice spacing, and nearest-neighbor hopping dynamics, is 
subject to a drive favoring movement along one axis of the square
lattice. 
The equilibrium (zero drive) transition to a phase with sublattice 
ordering, known to be continuous, shifts to lower density, and becomes 
discontinuous for large bias. In the ordered nonequilibrium steady 
state, both the particle and order-parameter densities are nonuniform, 
with a large fraction of the particles occupying a jammed strip 
oriented along the drive. The relaxation exhibits features
reminiscent of models of granular and glassy materials.
\end{abstract}


The study of simple nonequilibrium lattice
hopping models, such as the asymmetric exclusion process (ASEP), has
blossomed over the past decade, motivated by studies of traffic and of
granular matter \cite{krugprl,derrida,stinchcombe,wolf,schutz,nagel,kadanoff}.  
In parallel there is continuing interest in {\it driven
diffusive systems} (DDS): lattice-gases with biased 
hopping \cite{kls,ptcp17,marro,antal},
originally proposed as models of fast ionic 
conductors (FIC) \cite{fic}.
In the most widely studied case of DDS with attractive interactions,
application of a driving field E (favoring hopping along one of the
principal lattice directions) causes the interface between phases to
orient along the field.  Since the critical temperature 
increases with E, the drive may be said to favor ordering.

In DDS with {\it repulsive} 
interactions, which would appear to be more pertinent to ionic
conduction \cite{note1}, one expects the driving field to
suppress ordering (preferential occupation of
one sublattice).  Indeed, the original field
theoretic, simulation, mean-field analyses
\cite{leung,rddsmft}, showed that antiferromagnetic order
is destroyed generically for a drive $E\! >\! 2J$,
$J$ being the magnitude of the nearest-neighbor interaction.
More recent simulations \cite{szabo94,szabo96}, 
using larger lattices, strongly suggest
that global antiferromagnetic order is destroyed by {\it any} drive, 
however small.  The observation that FIC materials
exhibit phase transitions, even in the presence of repulsive
short-range interactions, and growing interest in
transport in particle systems, motivate 
investigation of a driven system with {\it hard-core} repulsion.

In this Letter I describe the case $J \!\to \!\infty$, i.e., a lattice gas
with occupancy of nearest-neighbor sites excluded (NNE).
(The distance between any pair of particles must be $ > 1$
lattice spacing; there are no other interactions.)
Equilibrium properties of this lattice analog of the hard-sphere
fluid were studied via series expansion methods in the 1960's,
leading to the conclusion that (on bipartite lattices) the system 
undergoes a continuous
phase transition at a critical density $\rho_c$, 
to a state with preferential occupancy of one sublattice;  
$\rho_c \simeq 0.37$ on the square lattice \cite{runnels,gf,ree}.
(The chemical potential is the temperaturelike parameter for 
this entropy-driven phase transition, which appears to fall
in the Ising universality class \cite{gf}.)  
To study the driven system, I adopt 
nearest-neighbor hopping dynamics.  Each particle has an
intrinsic rate of 1/4 for hopping in the $\pm y$ directions,
and of $p/2$ and $(1-p)/2$ in the $+x$ and $-x$ directions,
respectively.  Thus $p-1/2$ represents the deviation from
equilibrium, with $p=1$ corresponding to $E \to \infty$ in
standard DDS.  (Note however that in DDS the driving field
dominates the interparticle interaction when $E \gg J$, whereas
in the present case hard-core repulsion takes precedence over the drive.)

I study the driven NNE lattice gas on a square lattice of $L^2$
sites, using periodic boundary conditions in both 
directions.  Initially $N$ particles are thrown at random
onto the lattice, respecting the NNE condition;  
this yields a homogeneous,
disordered initial configuration.
[The random sequential adsorption (RSA) or ``parking"
process used to generate
the initial state eventually jams, at a mean density of
0.3641 \cite{rsa}.  In practice, one can reach densities
up to about 0.38 for $L=100$.]
In the dynamics a randomly selected
particle is assigned a hopping
direction according to the rates given above, and the new
position is accepted subject to the NNE condition.
$N$ such attempted moves define one time unit.
After a transient, whose duration depends on $p$, $L$,
and the density $\rho = N/L^2$, the system reaches a steady state
in which the current density $j$ (the net flux of 
particles along the $+x$ direction per site and unit time), and other 
macroscopic properties fluctuate about stationary values.
I report results of extensive simulations at $p=1$, 0.75, and 0.6;
some studies at $p=1/2$ were also performed to facilitate comparison
with equilibrium.

The stationary current density for $p\!=\!1$ is shown in Fig. 1.
For $\rho \leq 0.25$ there is little
dependence on system size; for the lower range of densities 
($\rho \leq 0.1$) the simulation data are in good agreement
with a four-site cluster mean-field calculation.  But at
higher densities ($\rho = 0.272$, 0.265, and 0.264 for $L=100$,
150, and 200, respectively), $j$ suddenly jumps to a lower value, and 
then continues to decrease smoothly with density.   These
data suggest a discontinuous phase transition 
(in the $L \to \infty$ limit) at a density
$\rho_c \simeq 0.263$, far below that of the equilibrium
critical point.
Before analyzing the evidence for this transition in greater detail,
some observations are in order.

First, for $p=1$ the system exhibits {\it jammed states:} absorbing
configurations in which each particle is blocked from moving by 
its neighbors (thus $j \equiv 0$).  While such configurations exist even at
very low densities (for example, an array
of particles occupying second-neighbor sites, forming
a rectangle spanning the system),
jammed states are not encountered in simulations
for densities less than about 0.32, for $L\geq 150$.  
In particular, jamming is
never observed in the vicinity of the transition to the
ordered phase.   Jamming occurs readily for $\rho \geq 0.34$;
in other words, the drive induces jamming at densities 
well below the RSA limit.  The system can jam for $p<1$
as well: while motion against the drive is possible in this case,
it appears that particles queue up to form dense configurations
whose lifetime, while finite, grows exponentially with system size.
For $p=0.75$ and $L=100$, jamming occurs at densities above about 0.37.

Jammed configurations are familiar from the {\it two species}
driven lattice gas, with the two kinds of particle driven
in opposite directions (and site exclusion the only 
interaction) \cite{schmitt92,ptcp17}.  In that case jammed 
states form rather more readily than in the NNE lattice gas
(for large driving fields the required density is $\approx 0.25$), and
are associated with a (possibly discontinuous) phase transition.
The existence of a sharp
jamming transition in the driven NNE system, of relevance to 
models of granular flow, will be addressed in future work.

Second, the two- and four-site cluster mean-field
theories (MFT) are so highly constrained by the
NNE condition that the cluster probabilities are {\it independent} 
of the drive $p$.  Pair MFT predicts
a continuous transition to a sublattice-ordered state at $\rho_c = 1/4$,
while the four-site approximation shown in Fig. 1
gives $\rho_c = 0.2696$.  But since in equilibrium $\rho_c \simeq 0.37$,
the fact that the kink in the MFT curve falls near
the discontinuous transition should be seen as purely fortuitous.  
Larger clusters, or a spatially extended MFT will be needed to study 
the driven NNE lattice gas.

The simulations extend to a maximum time of from 
$\sim 3 \times 10^6 $ to $\sim 6 \times 10^7$ steps,
depending on the time required to reach the stationary state.
Data for the stationary current and order parameter densities are 
obtained from {\it histograms} for the corresponding quantities,
in the final (stationary) stage.
Near the transition, at $p=1$ and 0.75, the histograms are bimodal 
(see Fig. 2).

In equilibrium, the phase transition in the NNE lattice gas is
signaled by a nonzero ``antiferromagnetic"
order parameter $\phi \!=\! |\rho_A \!-\! \rho_B|$, where $\rho_i = N_i/L^2$, 
with $N_i$ the number of particles in sublattice $i$.
Fig. 2 shows the evolution of $j(t)$ and $\phi(t)$ in a
single trial near $\rho_c$: the order parameter
increases suddenly, after a long waiting time, and then jumps
between high and low values; the current mirrors
the changes in the order parameter.
(The time to the onset of ordering varies from trial to trial,
within the range $10^5$ - $3 \times 10^6$, near $\rho_c$,
for $p=1$ and $L=150$.)

Fig. 3 shows the stationary order parameter as a function of 
density.  The phase transition is clearly discontinuous for
$p \geq 0.75$, and continuous for $p=0.6$ and 0.5 (equilibrium).  
In fact, for $p=0.6$ the system exhibits the hallmarks of a
critical point: the variation of the order parameter with $\rho$
becomes sharper with increasing system size;
the current varies smoothly with density (Fig. 1, inset); 
the histograms are unimodal, becoming very broad near the
transition, as reflected in a diverging variance of the order parameter
(see the lower inset of Fig. 3).  
There is presumably a {\it tricritical
point} at some drive $p_t$ between 0.6 and 0.75; determining the
precise value will require more extensive studies of larger systems.  
The transition from the disordered, high-conductivity state to
the ordered, low-conductivity state occurs at lower density,
the larger is $p$.  Thus we may cross the
phase boundary at fixed density by augmenting $p$ (Fig. 3, upper inset); 
the current, paradoxically, falls sharply in response to an increased drive!

The mechanism of the phase transition becomes clearer when we
examine particle configurations.  Fig. 4, a typical
stationary ordered configuration for $p=1$, reveals a strongly  
nonuniform distribution: a high-density strip has formed parallel
to the drive.  Within the strip, all particles occupy
a single sublattice, and their movement is blocked. 
(Comparison of configurations
at different times shows that while there are changes on the fringe,
the interior of the strip is frozen on time scales of at least 
$10^4$ steps.)  The low-density region outside the strip shows
no sublattice ordering, and harbors all the mobile
particles. 

For $p=1$ and 0.75, the strip often contains a central core of
maximum density, while the surrounding particles tend to 
fall along diagonal ``branches" terminating on the core, 
forming a herringbone pattern, or, as it were, 
an arrow pointing along the drive.  This suggests that the core
forms first, and that particles are subsequently trapped along
the branches.  The growth of the latter is limited by  
reduction of the density outside the strip.  For $p=0.6$
diagonal branches are again visible, but they are broader
(5-10 lattice spacings).  A very simple model of branch dynamics
(via addition or loss of particles at the tip outside the core, 
using a two-site MFT) yields growth for densities 
greater than 0.19, 0.26, and 0.29 in the neighborhood of the tip,
for $p=1$, 0.75, and 0.6, respectively.  (For smaller densities
the tip shrinks.)

Fig. 5 shows typical density and order-parameter profiles 
perpendicular to the drive.
The marked variation in the density
suggests that a second order parameter (the difference between
the maximium and minimum densities, $\Delta \rho =\rho_{max}-\rho_{min}$),
can be associated with the phase transition.  
This density difference is {\it the} order parameter 
in standard DDS (attractive NN interactions), but in that case
the equilibrium system also shows phase separation, whereas the 
equilibrium NNE lattice gas does not have high- and
low-density phases.  In the NNE system a nonzero 
$\Delta \rho$ is a uniquely nonequilibrium effect.

For $p=1$, ordered-state density profiles have 
$\rho_{max} \simeq 0.35$, essentially independent of $\rho$,
while $\rho_{min}$ generally falls the range 0.18 - 0.20,
again without a systematic $\rho$-dependence.  
For $p>1/2$, but less than unity, ordered configurations
again show a high-density strip coexisting with a low-density
region having a nonzero current and no sublattice ordering.
For $p=0.75$ I find $\rho_{max} \simeq 0.383$, while for
$p=0.6$, $\rho_{max} \simeq 0.40$. $\rho_{min}$ also 
increases with decreasing $p$: $\rho_{min} \simeq 0.19$,
0.253, and 0.285 for $p=1$, 0.75, and 0.6, resp., in reasonable
agreement with the branch-growth model mentioned above.
(MFT and numerical details will be reported elsewhere \cite{tobe}.)

Recall that for $p=1$, jamming occurs readily
for $\rho \simeq 0.34$, close to the density $\rho_{max}$ of the
jammed strip in the ordered state.  
Similarly, for $p=0.75$ the density in the strip is about 0.38,
close to the global jamming density of 0.37 or so.
(For $p=0.6$ the jamming density, presumably around 0.40, is beyond the 
density that can be prepared via RSA.)
The strip, then, appears
to represent an instability to {\it local} jamming in a system whose 
density lies below that needed for global jamming.
It is well known that driven particle systems such as the ASEP 
can exhibit a {\it shock}, i.e., a discontinuous density profile,
which back-propagates (in the direction opposite the drive) if the
current in the high-density region is smaller than in the low-density
region \cite{schutz}.  It is reasonable to suppose that such a shock forms 
in the present system due to a density fluctuation (perhaps as a result of
hopping perpendicular to the drive), and that its `tail' may then grow 
until it reaches the `head,' yielding a high-density ring which becomes
the core of the jammed strip.

Once it has formed, additional particles cannot readily
enter the jammed region, and so $\rho_{max}$ does not vary with $\rho$.
As the overall density $\rho$ increases, $\phi$ grows (and $j$
decreases) as a result of expansion of the jammed region. 
Eventually it fills the system; globally jammed configurations have
perfect sublattice ordering ($\phi \!=\! \rho$), and a uniform
density profile, except for empty triangular or diamond-shaped regions that
sometimes appear. 

I turn now to some preliminary results on dynamics. 
The apparent waiting time, $\tau$, to the onset 
of the stationary state typically falls in the range 10$^4$-10$^5$
for $\rho < \rho_c$.  At the transition it shows a
sharp maximum, 1 - 3 orders of magnitude above the pre-transition
value, and then falls off gradually ($\tau(\rho_c) \approx 2 \times 10^6$
for $p=1$ and $L=150$).  
Definitive results on relaxation times will require larger samples
than were used in this study.
Another interesting observation is the appearance, in some
realizations at $p=1$ and $\rho \geq 0.3$, 
of slow relaxation.  The order parameter 
grows linearly with $\ln t$ over a sizeable interval
(e.g., $6 \times 10^4$ - $3 \times 10^6$), before saturating.
This is reminiscent of the slow
compactification seen in granular materials.  
A model with purely excluded-volume interactions, but with rather
different boundary and driving conditions, has in fact been
found to reproduce slow (logarithmic) granular compaction \cite{caglioti}.
The present model may also be viewed as a `scalar' version of
the situation envisioned in Ref. \cite{cates}, in which application
of a shear stress provokes jamming (and thus rigidity) in a
granular medium.

The mean time $\tau_J$ to jamming appears to have
an exponential dependence on the density.  Studies at $p=1$
and $L=150$ give $\tau_J \simeq 10^4$
for $\rho \geq 0.34$, while for smaller densities $\tau_J$ grows
exponentially with $(0.34-\rho)$, increasing by more than two orders of
magnitude between $\rho = 0.34$ and 0.32; $\tau_J$ exceeds the
simulation time for $\rho < 0.31$.

Related to the first-order phase transition is the question of
hysteresis, and of the nucleation, growth, and decay of a
jammed strip.  The nonequilibrium critical behavior observed for
a smaller drive (e.g., $p=0.6$) is an important subject for
detailed study, since the nature of scaling in DDS remains
controversial \cite{ptcp17,marro}. Further issues
to be explored in future work are the effects of 
different initial configurations, of the aspect ratio
(in rectangular systems), of boundary conditions
(open along the drive, and/or reflecting perpendicular to it),
and systematic studies of
temporal and spatial correlations, and of finite-size effects.
Finally, it would be very useful to develop continuum descriptions
of this system, be they stochastic
(Langevin-like, starting from a suitable time-dependent
Landau-Ginzburg formulation), or deterministic (hydrodynamic,
starting perhaps from a kinetic theory of the lattice model).

In summary, I have shown that the phase transition in the 
lattice gas with nearest-neighbor exclusion persists under
a drive, and turns discontinuous for a sufficiently large bias.
Ordering is attended by the qualitatively new phenomenon (for a 
system with repulsive interactions) of phase segregation, which appears 
to result from an instability to formation of jammed regions.
As in the attractive DDS, certain features of the ordered state
can be understood on the basis of dynamic stability, as opposed to
interactions or free-energy considerations.
In the context of ASEP-like models, the results show that a
two-dimensional system is capable of exhibiting a {\it bulk}
phase transition, whereas the corresponding one-dimensional system
is expected to show only {\it boundary-induced} transitions.
The driven NNE lattice gas displays a surprisingly rich variety
of behaviors for its simplicity, and may be of relevance to
experiments on ionic conductors.
Certain aspects of its behavior may
yield insights into glassy and granular relaxation.
\vspace{2em}

\noindent {\bf Acknowledgements}
\vspace{1em}

\noindent 
I thank Robin Stinchcombe, Gyorgy Szab\'o, and Miguel Angel Mu\~noz 
for helpful discussions.
This work was supported in part by CNPq and CAPES.
\vspace{1em}

\noindent \vspace{1em}
 
\noindent $^*${\small electronic address: dickman@fisica.ufmg.br } \\

\newpage

\newpage

\noindent FIGURE CAPTIONS
\vspace{1em}

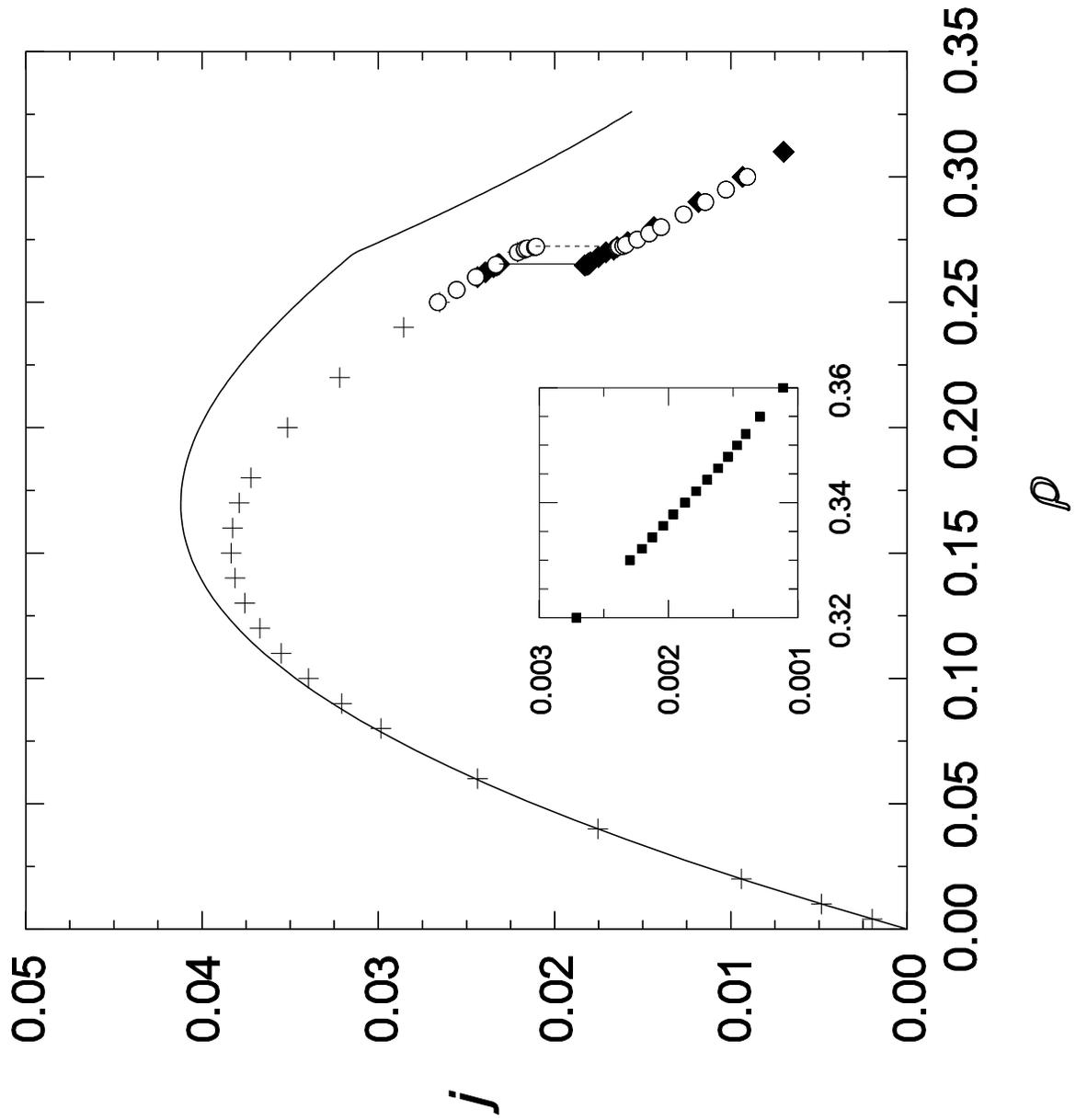
\noindent FIG. 1.  
Stationary current density versus particle density
for $p=1$ and $L=50$ (+), $L=100$ ($\circ$),
and $L=150$ (diamonds).  Vertical lines indicate the
transition at $L=100$ (dotted) and $L=150$ (solid).
Solid curve: four-site MFT.  Inset: $j$ versus $\rho$
for $p=0.6$ in the region of the transition, $L=100$.
\vspace{1em}

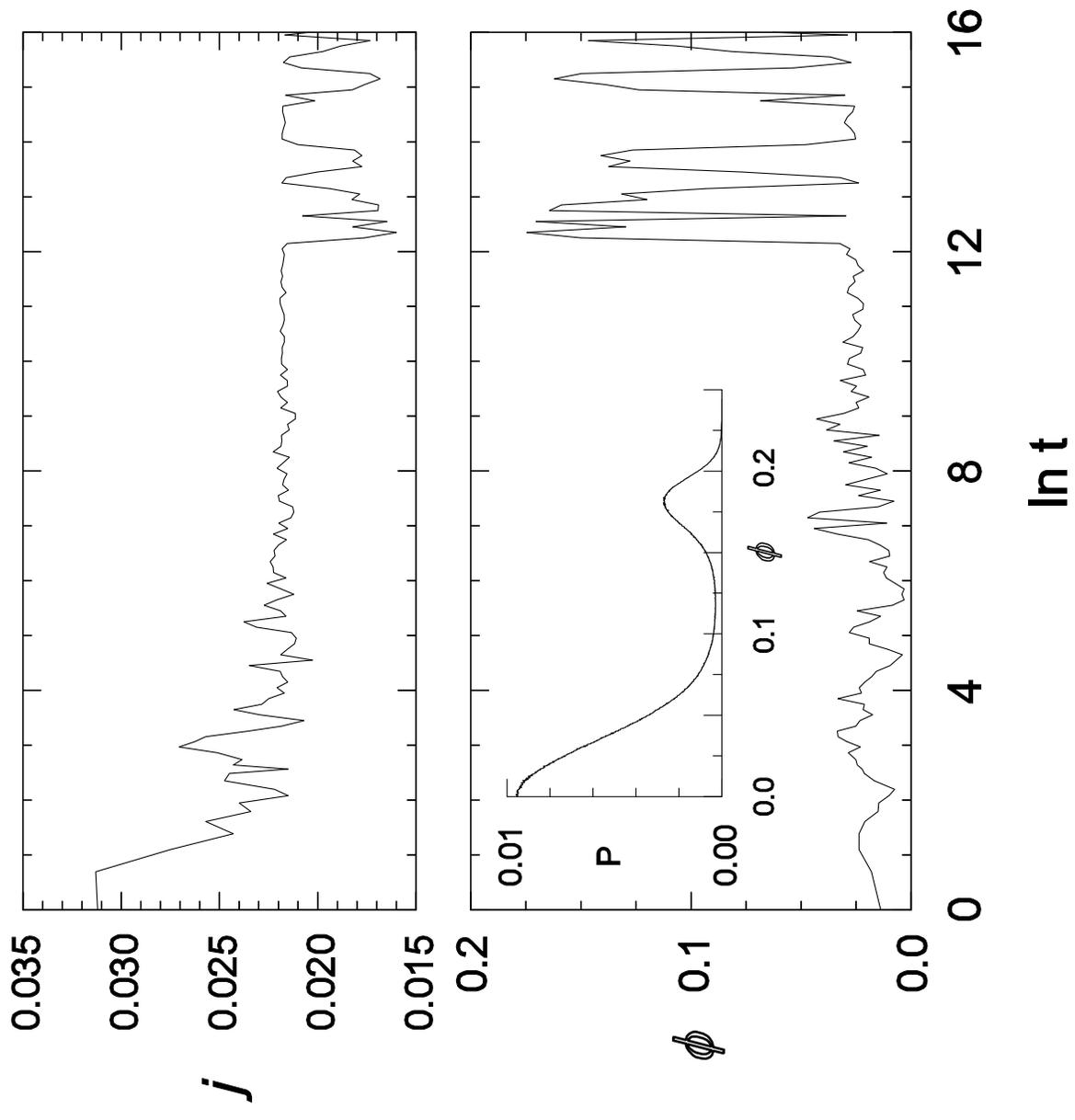
\noindent FIG. 2. Order parameter (lower panel) and current
density (upper) in a single trial at $p=1$, $\rho=0.2715$
and $L=100$.  Inset: order-parameter histogram for the same
run, accumulated for $t > 3.3\times 10^6$.
\vspace{1em}

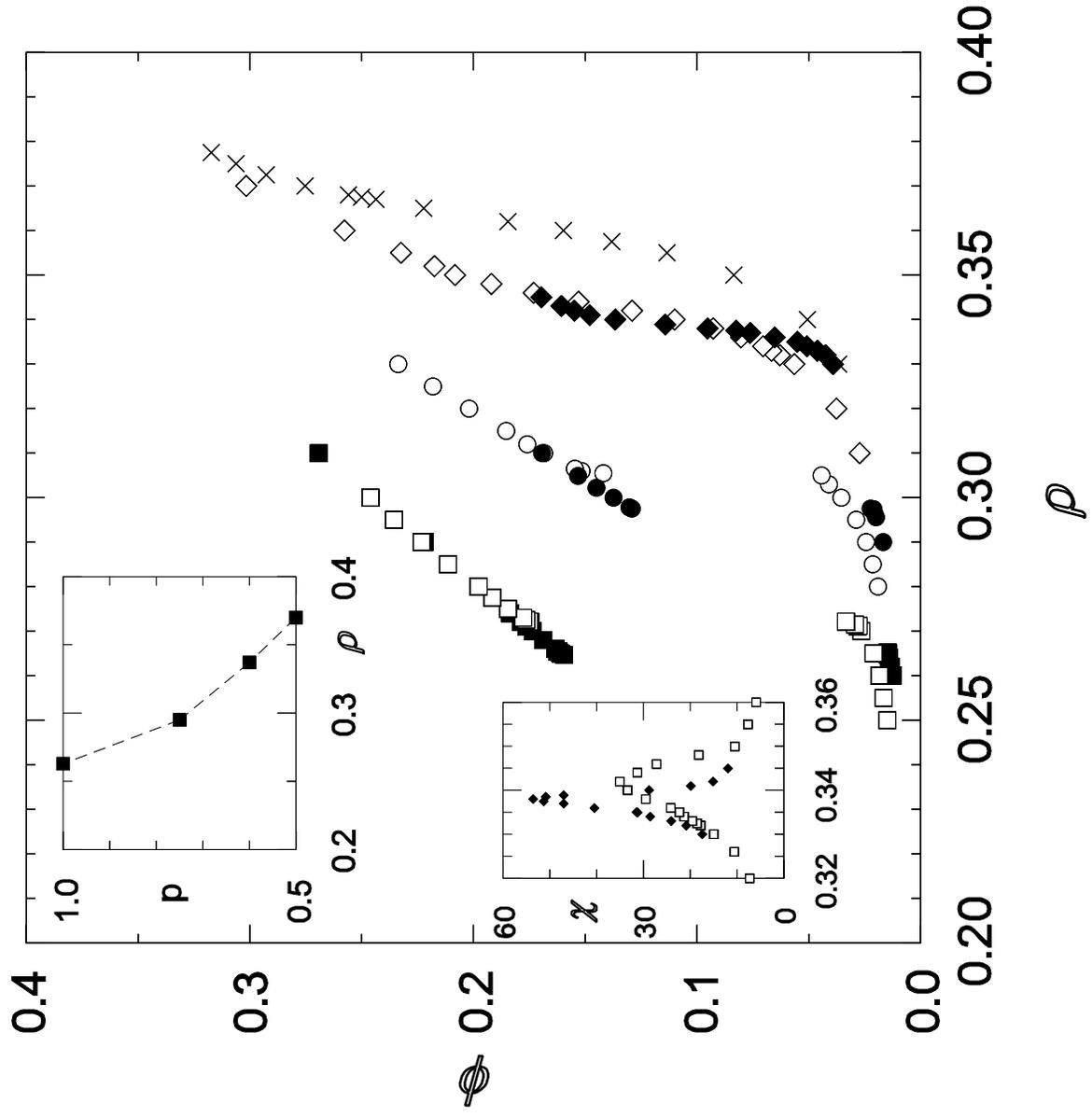
\noindent FIG. 3. Stationary order parameter versus density for $p=1$ 
(squares), $p=0.75$ (circles), and $p=0.6$ (diamonds).  Open symbols
are for $L=100$, filled, $L=150$.  Crosses represent the
equilibrium system ($p=1/2$) with $L=100$.
Upper inset: estimated phase boundary in the $\rho - p$ plane. 
Lower inset: $\chi \equiv L^2 {\mbox var}(\phi)$ versus density
for $p=0.6$ and $L=100$ (open symbols) and 150 (filled symbols).
\vspace{1em}

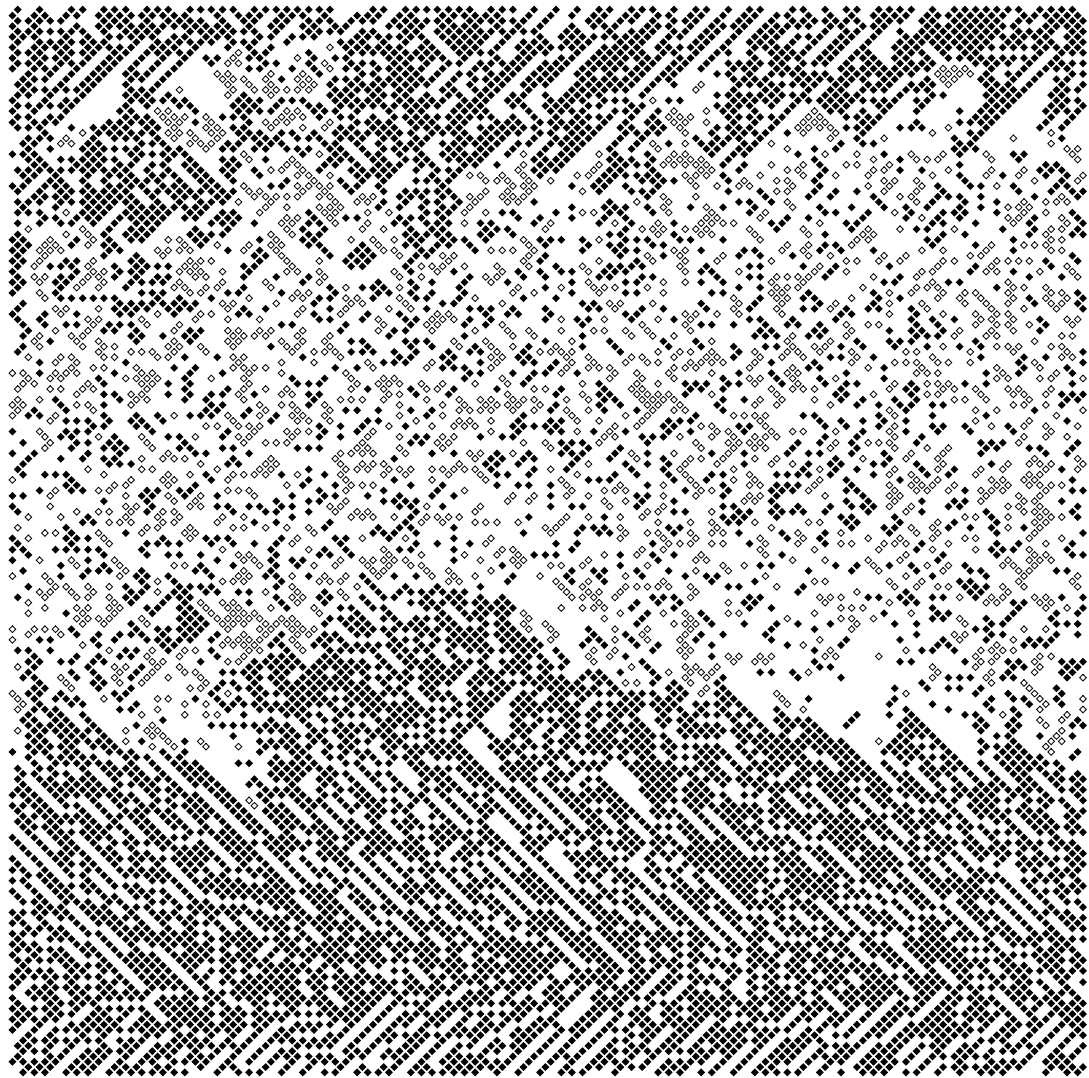
\noindent FIG. 4. 
Snapshot of particle configuration in stationary state, $p=1$,
$\rho = 0.267$, $L=200$.  Filled and open symbols represent
particles on different sublattices.  Driving field toward the
right.
\vspace{1em}

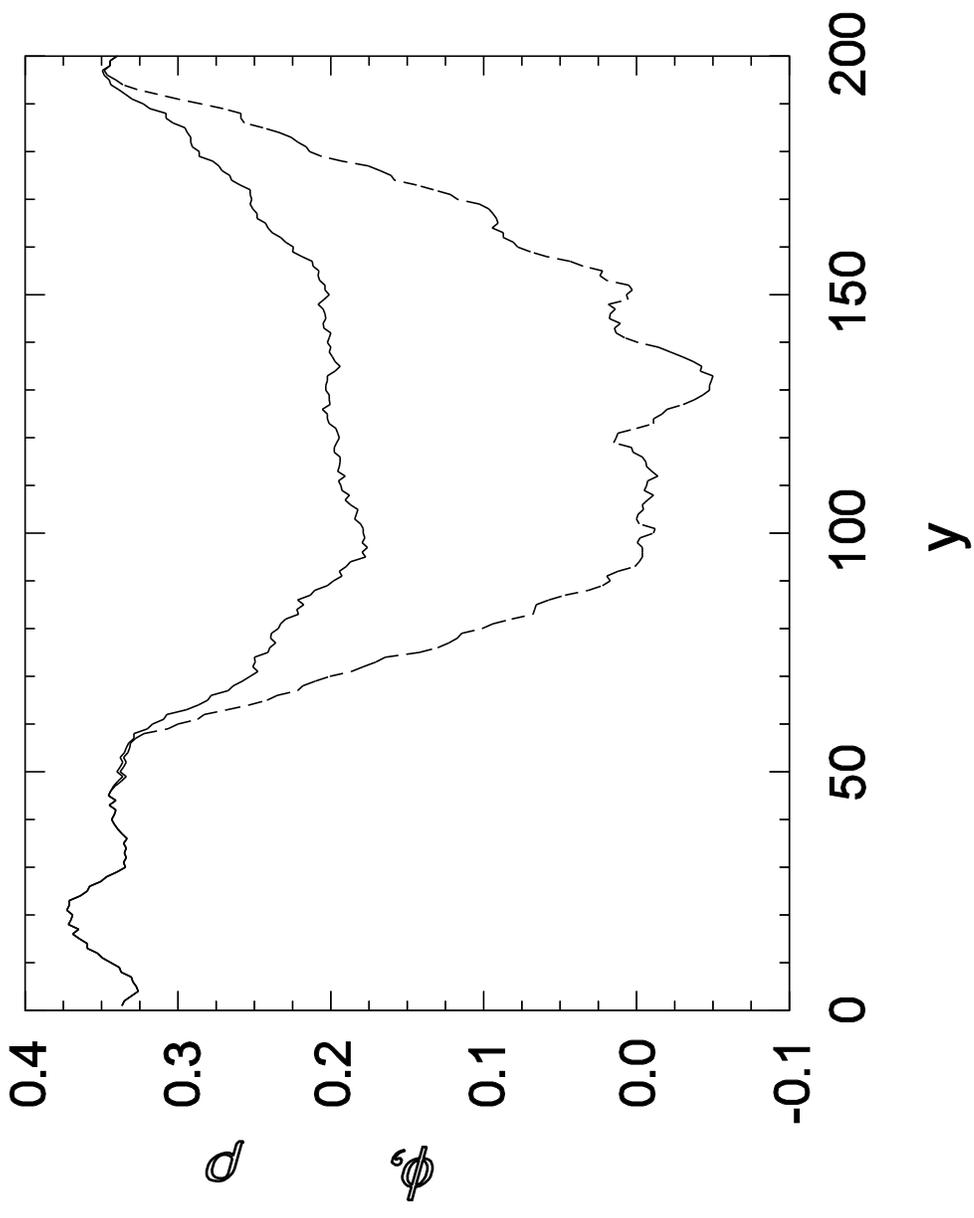
\noindent FIG. 5. 
Density (solid line) and order-parameter (broken line) profiles
perpendicular to the field for the configuration shown in Fig. 4.
\vspace{1em}


\begin{thebibliography}{99}
\bibitem{note1}
	While the crystal lattice may mediate an effective attractive
	contribution to the interaction, excluded-volume and Coulombic
	forces should in general yield a net repulsion
	between ions of the same species, at short range.
\end{thebibliography}
\end{document}